\documentclass{article}
\usepackage[dvips]{graphicx}  

\begin{document}

\newcommand{\rum}{\rule{0.5pt}{0pt}}

\newcommand{\rub}{\rule{1pt}{0pt}}

\newcommand{\rim}{\rule{0.3pt}{0pt}}

\newcommand{\numtimes}{\mbox{\raisebox{1.5pt}{${\scriptscriptstyle \times}$}}}

\renewcommand{\refname}{References}


\begin{center}

{\Large\bf  Dynamical 3-Space: Alternative Explanation of the `Dark Matter Ring'  \rule{0pt}{13pt}}\par

\bigskip

Reginald T. Cahill \\ 

{\small\it School of Chemistry, Physics and Earth Sciences, Flinders University,

Adelaide 5001, Australia\rule{0pt}{13pt}}\\

\raisebox{-1pt}{\footnotesize E-mail: Reg.Cahill@flinders.edu.au}\par

\bigskip\smallskip

{\small\parbox{11cm}{%

NASA has claimed the discovery of a `Ring of Dark Matter' in the galaxy cluster CL 0024+17, see Jee M.J. {\it et al.} arXiv:0705.2171, based upon gravitational lensing data.  Here we show that the lensing can be given an alternative explanation that does not involve `dark matter'.  This explanation comes from the new dynamics of 3-space.  This dynamics involves two constant $G$ and $\alpha$ - the fine structure constant. This dynamics  has explained the bore hole anomaly, spiral galaxy flat rotation speeds, the masses of black holes in spherical galaxies,   gravitational light bending and lensing, all without invoking `dark matter', and also the supernova redshift data without the need for  `dark energy'.  

\rule[0pt]{0pt}{0pt}}}\bigskip

\end{center}

\section{Introduction}

Jee  {\it et al.} \cite{Ring} claim that the analysis of gravitational lensing data from the HST observations  of the galaxy cluster CL 0024+17 demonstrates the existence of a `dark matter ring'.  While the lensing is clearly evident, as an observable phenomenon, it does not follow that this must be caused by some undetected  form of matter, namely  the putative `dark matter'.   Here we show that the lensing can be given an alternative explanation that does not involve `dark matter'.  This explanation comes from the new dynamics of 3-space \cite{Book,Schrod,alpha,DM,galaxies}.  This dynamics involves two constant $G$ and $\alpha$ - the fine structure constant. This dynamics  has explained the bore hole anomaly, spiral galaxy flat rotation speeds, the masses of black holes in spherical galaxies,   gravitational light bending and lensing, all without invoking `dark matter'. The 3-space dynamics  also has a Hubble expanding 3-space solution that explains the supernova redshift data without the need for  `dark energy' \cite{DE}.   The issue is that the Newtonian theory of gravity \cite{Newton}, which was based upon observations of planetary motion in the solar system, missed a key dynamical effect that is not manifest in this system.  The consequences of this failure has been the  invoking of  the fix-ups  of `dark matter' and `dark energy'.  What is missing is the 3-space self-interaction effect. Experimental and observational data has shown that the coupling constant for this self-interaction is the fine structure constant, $\alpha \approx 1/137$, to within  measurement errors.
It is shown here that this 3-space self-interaction effect gives a direct explanation for the reported ring-like gravitational lensing effect. 

\begin{figure}[t]
\vspace{0mm}\,\parbox{93mm}{\includegraphics[width=90mm]{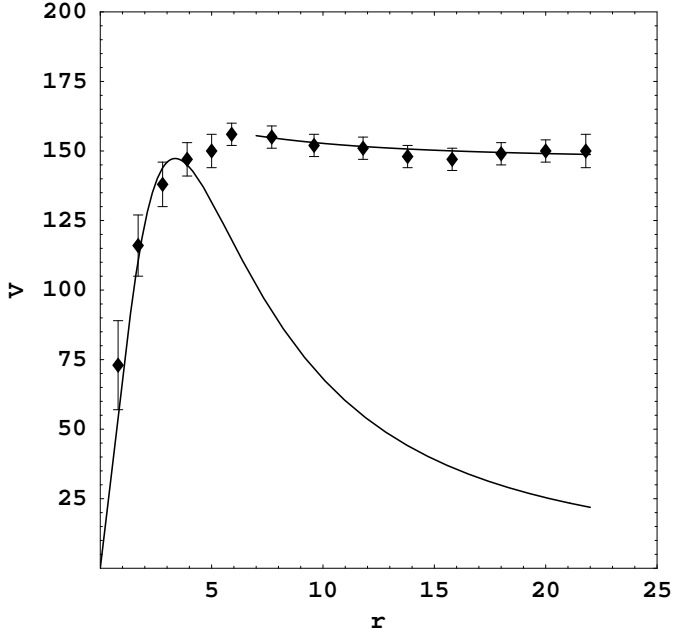}}\,
\parbox{70mm}{\caption{\small{  Data shows the non-Keplerian rotation-speed curve $v_O$ for the spiral galaxy NGC 3198 in km/s plotted
against radius in kpc/h. Lower curve is the rotation curve from the Newtonian theory  for an
exponential disk, which decreases asymptotically like $1/\sqrt{r}$. The upper curve shows the asymptotic form from
(\ref{eqn:vorbital}), with the decrease determined by the small value of $\alpha$.  This asymptotic form is caused by
the primordial black holes at the centres of spiral galaxies, and which play a critical role in their formation. The
spiral structure is caused by the rapid in-fall towards these primordial black holes.}}}
\label{fig:NGC3198}\end{figure}

\section{ 3-Space Dynamics}
As discussed elsewhere \cite{Book,DE} a deeper information- theoretic {\it Process Physics}   has an emergent  structured 3-space, where the 3-dimensionality is partly modelled at a phenomenological level by embedding the time- dependent structure  in an $E^3$ or $S^3$  embedding space. This embedding space is not real -  it serves to coordinatise the structured 3-space, that is,  to provide an abstract frame of reference. 
Assuming the simplest dynamical description for zero-vorticity spatial velocity field ${\bf v}({\bf r},t)$,  based upon covariant scalars we obtain at lowest order \cite{Book}.
\begin{equation}
\nabla.\left(\frac{\partial {\bf v} }{\partial t}+({\bf v}.{\bf \nabla}){\bf v}\right)
+\frac{\alpha}{8}\left((tr D)^2 -tr(D^2)\right)=
-4\pi G\rho
\label{eqn:E1}\end{equation}
\begin{equation}
\nabla\times{\bf v}={\bf 0}, \mbox{\ \ \ \ \ \ }D_{ij}=\frac{1}{2}\left(\frac{\partial v_i}{\partial x_j}+
\frac{\partial v_j}{\partial x_i}\right)
\label{eqn:E1b}\end{equation}
where $\rho({\bf r},t)$ is the matter and EM energy density expressed as an effective matter density. 
In Process Physics  quantum matter  are topological defects in the structured 3-spaces, but here it is sufficient to give a simple description in terms of an  effective density. 

We see that there are two constants $G$ and $ \alpha$. $G$ turns out  to be Newton's gravitational constant, and describes the rate of non-conservative flow of 3-space into matter, and $\alpha$  is revealed by experiment to be the fine structure constant.   Now the acceleration ${\bf a}$ of the dynamical patterns of 3-space is given by the Euler convective expression
\begin{eqnarray}
{\bf a}({\bf r},t)&=& \lim_{\Delta t \rightarrow 0}\frac{{\bf v}({\bf r}+{\bf v}({\bf r},t)\Delta t,t+\Delta
t)-{\bf v}({\bf r},t)}{\Delta t} \nonumber \\
&=&\frac{\partial {\bf v}}{\partial t}+({\bf v}.\nabla ){\bf v}
\label{eqn:E3}\end{eqnarray} 
and this appears in the first term in (\ref{eqn:E1}). As shown in \cite{Schrod} the acceleration of quantum matter  ${\bf g}$ is identical to this acceleration, apart from vorticity and relativistic effects, and so the gravitational acceleration of matter is also given by (\ref{eqn:E3}).
Eqn.(\ref{eqn:E1}) is highly non-linear, and indeed non-local. It exhibits a range of different phenomena, and as has been shown the $\alpha$ term is responsible for all those  effects attributed to the undetected and unnecessary `dark matter'.
For example, outside of a spherically symmetric distribution of matter,  of total mass $M$, we find that one solution of (\ref{eqn:E1}) is the velocity in-flow field  
\begin{equation}
{\bf v}({\bf r})=-\hat{{\bf r}}\sqrt{\frac{2GM(1+\frac{\alpha}{2}+..)}{r}}
\label{eqn:E4}\end{equation}
and then the the acceleration of (quantum) matter, from (\ref{eqn:E3}), induced by this in-flow is
\begin{equation}
{\bf g}({\bf r})=-\hat{{\bf r}}\frac{GM(1+\frac{\alpha}{2}+..)}{r^2}
\label{eqn:E5}\end{equation}
 which  is Newton's Inverse Square Law of 1687 \cite{Newton}, but with an effective  mass $M(1+\frac{\alpha}{2}+..)$ that is different from the actual mass $M$. 

In general because (\ref{eqn:E1}) is a scalar equation it is only applicable for vorticity-free flows $\nabla\times{\bf v}={\bf 0}$, for then we can write ${\bf v}=\nabla u$, and then (\ref{eqn:E1}) can always be solved to determine the time evolution of  $u({\bf r},t)$ given an initial form at some time  $t_0$.
The $\alpha$-dependent term in (\ref{eqn:E1})  and the matter acceleration effect, now also given by (\ref{eqn:E3}),   permits   (\ref{eqn:E1})   to be written in the form
\begin{equation}
\nabla.{\bf g}=-4\pi G\rho-4\pi G \rho_{DM},
\label{eqn:E7}\end{equation}
\begin{equation}
\rho_{DM}({\bf r},t)\equiv\frac{\alpha}{32\pi G}( (tr D)^2-tr(D^2)),  
\label{eqn:E7b}\end{equation}
which  is an effective `matter' density that would be required to mimic the
 $\alpha$-dependent spatial self-interaction dynamics.
 Then (\ref{eqn:E7}) is the differential form for Newton's law of gravity but with an additional non-matter effective matter density.  So we label this as $\rho_{DM}$ even though no matter is involved \cite{alpha,DM}. This effect has been shown to explain the so-called `dark matter' effect in spiral galaxies, bore hole $g$ anomalies, and the systematics of galactic black hole masses.    
 
 The spatial dynamics  is non-local.  Historically this was first noticed by Newton who called it action-at-a-distance. To see this we can write  (\ref{eqn:E1}) as an integro-differential equation
 \begin{equation}
 \frac{\partial {\bf v}}{\partial t}=-\nabla\left(\frac{{\bf v}^2}{2}\right)+G\!\!\int d^3r^\prime
 \frac{\rho_{DM}({\bf r}^\prime, t)+\rho({\bf r}^\prime, t)}{|{\bf r}-{\bf r^\prime}|^3}({\bf r}-{\bf r^\prime})
 \label{eqn:E8}\end{equation}

 This shows a high degree of non-locality and non-linearity, and in particular that the behaviour of both $\rho_{DM}$ and $\rho$ manifest at a distance irrespective of the dynamics of the intervening space. This non-local behaviour is analogous to that in quantum systems and may offer a resolution to the horizon problem. 

 \begin{figure}[t]
\vspace{0mm}\,\parbox{105mm}{\includegraphics[width=100mm]{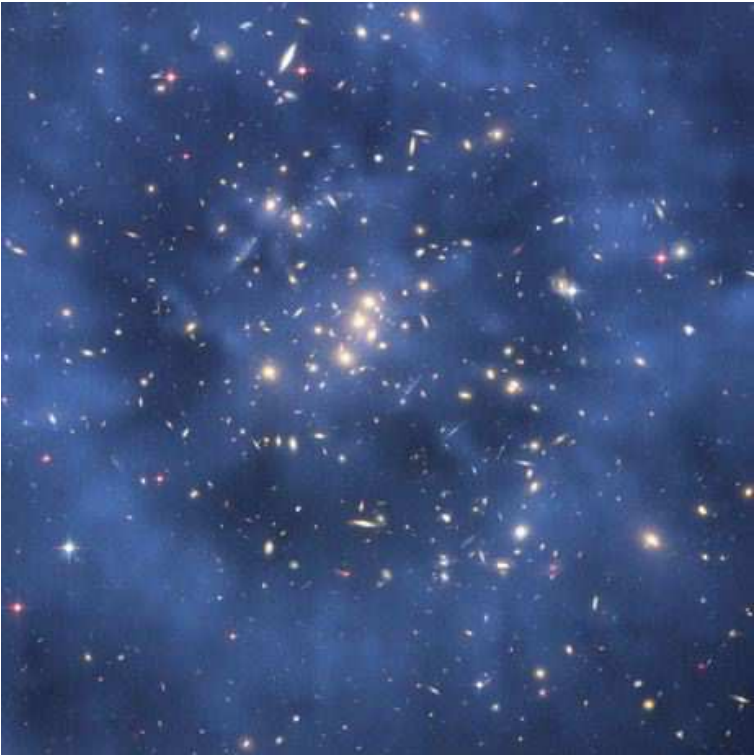}}\,
\parbox{65mm}{\caption{\small{   The `dark matter' density extracted by deconvolution of  the gravitational lensing data  for galaxy cluster CL 0024+17, see Jee M.J. {\it et al.} arXiv:0705.2171.  Picture credit: NASA, ESA, M.J. Jee and H.C.  Ford (John Hopkins University).  The `dark matter' density has been superimposed on a HST image of the cluster.  The axis of  `symmetry' is perpendicular to the planer of this image.  The gravitational lensing is caused by  two galaxy clusters that have undergone collision. It is claimed herein that the lensing is associated with the 3-space interaction of these two  `nearby'  galaxy clusters, and not by the fact that they had collided, as claimed in \cite{Ring}.  The effect it is claimed, herein, is caused by the spatial in-flows into the black holes within the galaxies.}}}
\label{fig:NASA}\end{figure}

\begin{figure}
\vspace{0mm}\,\parbox{105mm}{\includegraphics[width=100mm]{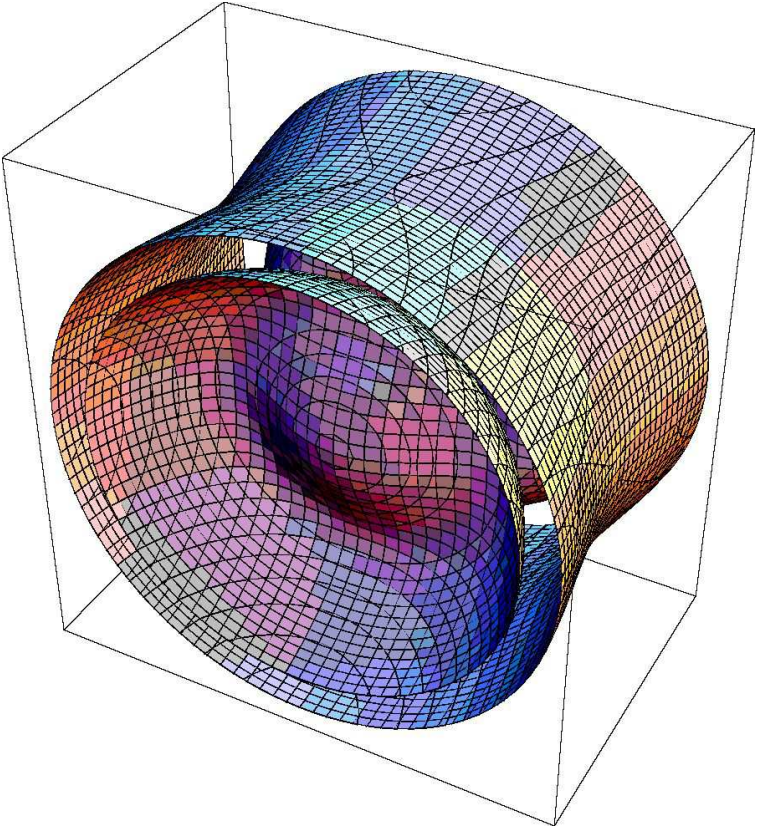}}\,
\parbox{65mm}{\caption{\small {Plot showing two constant-value surfaces of  $\Delta \rho_{DM}({\bf r })$ from (\ref{eqn:DeltaDM}). We have modelled the system with  two galaxies  located  on the axis of symmetry, but outside of the range of the plot.   This  plot shows the effects of the interfering spatial in-flows generating  an effective `dark matter' density, as a spatial self-interaction effect.  This `dark matter' density is that required to reproduce the gravitational acceleration if we used Newton's law of gravity. This phenomenon is caused by the $\alpha$-dependent dynamics in (\ref{eqn:E1}), essentially a quantum-space effect.  Viewed along the axis of symmetry this shell structure would appear as a ring-like structure, as seen in Fig.\ref{fig:NASA}.}}
\label{fig:ThreePlot}}\end{figure}

\begin{figure}[t]
\vspace{2mm}
\hspace{25mm}\includegraphics[scale=0.35]{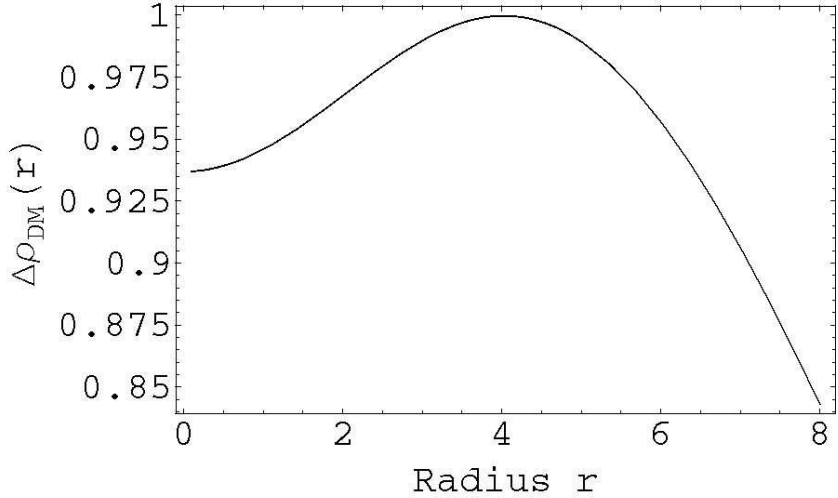}
\vspace{0mm}
\caption{\small {Plot of $\Delta \rho_{DM}({\bf r })$ from (\ref{eqn:DeltaDM}) in a radial direction from a midpoint on the axis joining the two galaxies.}
\label{fig:RadialPlot}}\end{figure}

\
 
\begin{figure}[t]
\vspace{2mm}
\hspace{30mm}\includegraphics[scale=0.3]{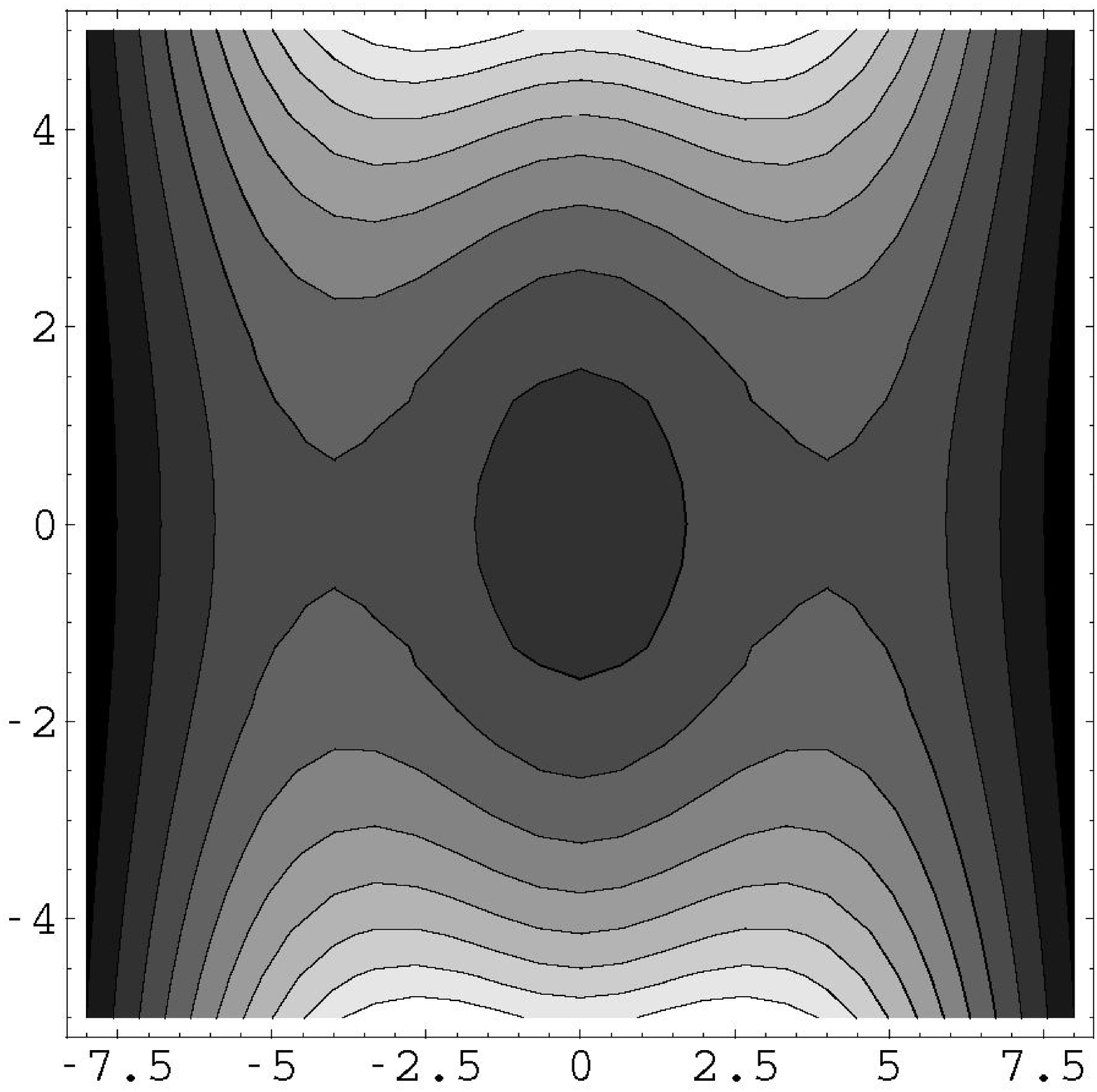}
\vspace{0mm}
\caption{\small {Plot of $\Delta \rho_{DM}({\bf r })$ from (\ref{eqn:DeltaDM}) in the plane containing the two galaxies. The two galaxies are located at +10 and -10, i.e above and below the vertical in this contour plot. This  plot shows the effects of the interfering in-flows.}
\label{fig:ContPlot}}\end{figure}

\subsection{Spiral Galaxy Rotation Anomaly}

Eqn  (\ref{eqn:E1}) gives also a direct explanation for the spiral galaxy rotation anomaly.    For a non-spherical system numerical solutions of (\ref{eqn:E1}) are required, but sufficiently far from the centre, where $\rho=0$,  we find an exact non-perturbative two-parameter class of analytic solutions
\begin{equation}
{\bf v}({\bf r}) =-\hat{\bf r} K\left(\frac{1}{r}+\frac{1}{R_s}\left(\frac{R_s}{r}  \right)^{\displaystyle{\frac{\alpha}{2}}}  \right)^{1/2}
\label{eqn:vexact}\end{equation}
where $K$ and $R_s$ are arbitrary constants in the $\rho=0$ region, but whose values are determined by matching to the solution in the matter region. Here $R_s$ characterises the length scale of the non-perturbative part of this expression,  and $K$ depends on $\alpha$, $G$ and details of the matter distribution.   From (\ref{eqn:E5})  and (\ref{eqn:vexact}) we obtain a replacement for  the Newtonian  `inverse square law' 
\begin{equation}
{\bf g}({\bf r})=-\hat{\bf r}\frac{K^2}{2} \left( \frac{1}{r^2}+\frac{\alpha}{2rR_s}\left(\frac{R_s}{r}\right)
^{\displaystyle{\frac{\alpha}{2}}} 
\right)
\label{eqn:gNew}\end{equation}
The 1st term, $1/r^2$, is the Newtonian part.  The 2nd term is caused by a `black hole' phenomenon that  (\ref{eqn:E1}) exhibits.  This manifests in different ways, from a minimal supermassive black holes, as seen in spherical star systems, from  globular clusters to spherical galaxies for which the black hole mass is predicted to be $M_{BH}=\alpha M/2$, as confirmed by the observational datas \cite{Book,alpha,DM,galaxies,newBH},  to primordial supermassive black holes as seen in spiral galaxies as described by (\ref{eqn:vexact}); here the matter spiral is caused by matter in-falling towards the primordial black hole.

The spatial-inflow phenomenon in (\ref{eqn:vexact})  is completely different from the putative `black holes' of General Relativity - the new `black holes' have an essentially  $1/r$ force law, up to $O(\alpha)$ corrections,  rather than the usual Newtonain and GR $1/r^2$ law.
  The centripetal acceleration  relation for circular orbits 
$v_O(r)=\sqrt{rg(r)}$  gives  a `universal rotation-speed curve'
\begin{equation}
v_O(r)=\frac{K}{2} \left( \frac{1}{r}+\frac{\alpha}{2R_s}\left(\frac{R_s}{r}\right)
^{\displaystyle{\frac{\alpha}{2}}} 
\right)^{1/2}
\label{eqn:vorbital}\end{equation}
 Because of the $\alpha$ dependent part this rotation-velocity curve  falls off extremely slowly with $r$, as is indeed observed for spiral galaxies. An example is shown in Fig.\ref{fig:NGC3198}. It was the inability of the  Newtonian  and Einsteinian gravity theories to explain these observations that led to the  notion of `dark matter'.

For the spatial flow in (\ref{eqn:vexact}) we may compute the effective dark matter density from (\ref{eqn:E7b})
 \begin{equation}
\tilde{\rho}_{DM}(r) = \frac{(1-\alpha)\alpha}{16\pi G}\frac{K^2}{R^3_s}\left(\frac{R_s}{r}  \right)^{2+\alpha/2} 
\label{eqn:DMDensity}\end{equation}
We see the standard $1/r^2$ behaviour. It should be noted  that the Newtonian component of  (\ref{eqn:vexact}) does not contribute, and that $\tilde{\rho}_{DM}({\bf r})$ is exactly zero in the limit $\alpha\rightarrow 0$.  So supermassive black holes and the spiral galaxy rotation anomaly are all $\alpha$-dynamics phenomena.

\subsection{Gravitational Lensing}
The spatial velocity field may be observed on the cosmological scale by means of the light bending and lensing effect. But first
we must   generalise the Maxwell equations so that the electric and magnetic  fields are excitations of the dynamical 3-space, and not of the embedding space: 
 \begin{equation}
\displaystyle{ \nabla \times {\bf E}=-\mu\left(\frac{\partial {\bf H}}{\partial t}+{\bf v.\nabla H}\right)},
 \mbox{\ \ \ }\displaystyle{\nabla.{\bf E}={\bf 0}},
\end{equation}
 \begin{equation} \displaystyle{ \nabla \times {\bf H}=\epsilon\left(\frac{\partial {\bf E}}{\partial t}+{\bf v.\nabla E}\right)} ,
\mbox{\ \ \  }\displaystyle{\nabla.{\bf H}={\bf 0}}\label{eqn:E18}\end{equation}
which was first suggested by Hertz in 1890, but with ${\bf v}$ being a constant vector field. As easily determined  the speed of EM radiation is $c=1/\sqrt{\epsilon\mu}$ wrt to the dynamical space, and not wt to the embedding space as in the original form of Maxwell's equations, and as light-speed  anisotropy experiment have indicated \cite{Book}. The time-dependent and inhomogeneous  velocity field causes the refraction of EM radiation. This can be computed by using the Fermat least-time approximation. Then the EM trajectory ${\bf r}(t)$ is determined by minimising  the elapsed travel time:
\begin{equation}
\tau=\int_{s_i}^{s_f}\frac{ds\displaystyle{|\frac{d{\bf r}}{ds}|}}{|c\hat{{\bf v}}_R(s)+{\bf v}(\bf{r}(s),t(s)|}
\label{eqn:path}\end{equation}
\begin{equation}
{\bf v}_R=\left(  \frac{d{\bf r}}{dt}-{\bf v}(\bf{r},t)\right)
\label{eqn:nvector}\end{equation}
by varying both ${\bf r}(s)$ and $t(s)$, finally giving ${\bf r}(t)$. Here $s$ is a path parameter, and ${\bf v}_R$ is a 3-space tangent vector for the path.
As an example, the  in-flow in (\ref{eqn:E4}), which is applicable to light bending by the sun, gives  the angle of deflection 
\begin{equation}
\delta=2\frac{v^2}{c^2}=\frac{4GM(1+\frac{\alpha}{2}+..)}{c^2d}+...
\label{eqn:E19}\end{equation}
where $v$ is the in-flow speed at distance $d$  and $d$ is the impact parameter. This agrees with the GR result except for the $\alpha$ correction.  Hence the  observed deflection of $8.4\times10^{-6}$ radians is actually a measure of the in-flow speed at the sun's surface, and that gives $v=615$km/s.   These generalised Maxwell equations also predict gravitational lensing produced by the large in-flows from (\ref{eqn:vexact}) associated with the new `black holes' in galaxies.  So again this effect permits the direct observation of the these  black hole effects with their non inverse-square-law accelerations. 

\section{Galaxy Cluster Lensing}

It is straightforward to analyse the gravitational lensing predicted by  a galaxy cluster, with the data from CL 0024+17 of particular interest. However rather than compute the actual lensing images, we shall compute the `dark matter' effective density from (\ref{eqn:E7b}), and compare that with the putative `dark matter' density extracted from the actual lensing data in \cite{Ring} .  To that end we need to solve (\ref{eqn:E1}) for two reasonably close galaxies, located at positions ${\bf R}$ and $-{\bf R}$.  Here we look for a perturbative modification of the 3-space inf-lows when the two galaxies are nearby. We take the velocity field in 1st approximation to be the superposition 
\begin{equation}
{\bf v}({\bf r})\approx\tilde{{\bf v}}({\bf r}-{\bf R})+\tilde{{\bf v}}({\bf r}+{\bf R})
\label{eqn:vsum}\end{equation}
where the tilde denotes single galaxy velocity field in (\ref{eqn:vexact}).   Substituting this in (\ref{eqn:E1}) will then generate an improved  solution, keeping in mind that  (\ref{eqn:E1}) is non-linear, and so this superposition cannot be exact. Indeed it is the non-linearity effect which it is claimed herein is responsible for the ring-like structure reported in \cite{Ring}.
Substituting  (\ref{eqn:vsum}) in (\ref{eqn:E7b}) we may compute the change in the effective `dark matter' density  caused by the two galaxies interfering with the in-flow into each separately, i.e.
\begin{equation}
\Delta \rho_{DM}({\bf r })=\rho_{DM}({\bf r}) -\tilde{\rho}_{DM}({\bf r}-{\bf R})-\tilde{\rho}_{DM}({\bf r}+{\bf R})
\label{eqn:DeltaDM}\end{equation}
$\tilde{\rho}_{DM}({\bf r}\pm{\bf R})$ are the the effective `dark matter' densities for one isolated galaxy in (\ref{eqn:DMDensity}). Several graphical representations of $\Delta \rho_{DM}({\bf r })$ are given in Figs. \ref{fig:ThreePlot}, \ref{fig:RadialPlot} and \ref{fig:ContPlot}.  We seen that viewed along the line of the two galaxies the change in the effective `dark matter' density has the form of a ring, in particular one should compare  the predicted effective `dark matter' density in Fig.\ref{fig:ThreePlot}  with that found by deconvoluting the gravitaitaional lensing data shown in shown  Fig.\ref{fig:NASA}.

\section{Conclusions}
We have shown that the dynamical 3-space theory gives a direct account of the observed gravitational lensing caused by  two  galaxy clusters, which had previously  collided, but that the ring-like structure is not related to that collision, contrary to the claims in \cite{Ring}. The distinctive lensing effect is caused by interference between the two spatial in-flows, resulting in EM  refraction which appears to be caused by the presence of   `matter' having the  form of a ringed-shell structure, exactly comparable to the observed effect. This demonstrates yet another  success of the new dynamical theory of 3-space, which like the bore hole, black hole and spiral galaxy rotation effects all reveal the dynamical consequences of the $\alpha$-dependent term in (\ref{eqn:E1}).  This amounts to a totally different understanding of the nature of space, and a completely different account of gravity. As shown in \cite{Schrod} gravity is a quantum effect where the quantum waves are refracted by the 3-space, and that analysis also gave a first derivation of the equivalence principle.  We see again that `dark matter' and `dark energy' are spurious concepts required only because Newtonian gravity, and {\it ipso facto} GR,  lacks  fundamental processes of a dynamical 3-space - they are merely {\it ad hoc} fix-ups.  We have shown elsewhere \cite{newBH} that from  (\ref{eqn:E1}) and the generalised Dirac equation  that  a curved  spacetime formalism may be introduced that permits the determination of the quantum matter geodesics, but that in general the spacetime metric does not satisfy the Hilbert-Einstein equations, as of course GR lacks the  $\alpha$-dependent dynamics. This induced spacetime has no ontological significance.  At a deeper level the occurrence of $\alpha$ in (\ref{eqn:E1}) suggests that 3-space is actually a quantum system, and that (\ref{eqn:E1}) is merely a phenomenological description of that at the `classical' level. In which case the $\alpha$-dependent dynamics amounts to the detection of quantum space and quantum gravity effects, although clearly not of the form suggested by the quantisation of GR.


\begin{thebibliography}{99}

\bibitem{Ring}  Jee M.J. {\it et al.} {\it Discovery of a Ring-Like Dark Matter Structure in the Core of the Galaxy Cluster CL 0024+17,} arXiv:0705.2171,   to be published in {\it  The Astrophysical Journal}.

\bibitem{Book} Cahill  R.T. {\it Process Physics: From Information Theory to Quantum Space
       and Matter},  Nova Science Pub., New York, 2005.
        \bibitem{Schrod} Cahill R.T. {\it  Dynamical  Fractal  3-Space and the Generalised Schr\"{o}dinger  
Equation: Equivalence Principle and  Vorticity Effects},   {\it Progress in Physics},  {\bf 1}, 27-34, 2006.
\bibitem{alpha}  Cahill R.T.   {\it Gravity, `Dark Matter' and the Fine Structure Constant}, {\it Apeiron}, {\bf
12}(2), 144-177, 2005.
 \bibitem{DM}   Cahill R.T.   {\it  `Dark Matter' as a Quantum Foam In-flow Effect}, in {\it
Trends in Dark Matter Research},  96-140,  ed. J. Val Blain , Nova Science Pub., New York, 2005.    
\bibitem{galaxies}  Cahill   R.T. {\it Black Holes in Elliptical and Spiral Galaxies and in 
Globular Clusters}, {\it Progress in Physics}, {\bf 3}, 51-56, 2005.
\bibitem{newBH} Cahill  R.T. {\it Black Holes and Quantum Theory: The Fine Structure Constant Connection},  {\it Progress in Physics}, {\bf 4}, 44-50, 2006.

\bibitem{DE} Cahill  R.T. {\it Dynamical 3-Space:  Supernova  and the Hubble Expansion -  Older Universe and End of  Dark Energy,}  arXiv:0705.1569.

\bibitem{Newton}  Newton I.  {\it Philosophiae Naturalis Principia Mathematica}, 1687.

\end{thebibliography}
\end{document}